\documentclass[twocolumn,pra,aps,superscriptaddress]{revtex4}

\usepackage{epsfig}
\usepackage{graphicx}
\usepackage{dcolumn}
\usepackage{bm} 
\usepackage{epstopdf}
\usepackage{amsmath}

\newcommand{\beq}{\begin{equation}}
\newcommand{\eeq}{\end{equation}}
\newcommand{\beqa}{\begin{eqnarray}}
\newcommand{\eeqa}{\end{eqnarray}}

\newcommand{\om}{\omega}

\def\jpb#1{{ J.\ Phys.\ B} {\bf#1}}

\def\natphys#1{{ Nature\ Phys.\ } {\bf#1}}

\def\njp#1{{ New J.\ Phys.\ } {\bf#1}}
\def\pra#1{{ Phys.\ Rev. A\/} {\bf#1}}

\def\prl#1{{ Phys.\ Rev.\ Lett.} {\bf#1}}
\def\pr#1{{ Phys.\ Rev.\/} {\bf#1}}

\def\rmp#1{{ Rev.\ Mod.\ Phys.} {\bf#1}}

\def\nat#1{{ Nature} {\bf#1}}

\begin{document}

\title{Virtual Detector Theory of Free-Particle Wave Packets}

\author{Xu Wang}
\email{xwang@gscaep.ac.cn}
\affiliation{Graduate School, China Academy of Engineering Physics, Beijing 100193, China}

\date{\today}

\begin{abstract} We apply a recently emerging ``virtual detector" method to general free-particle wave packets. We discuss what is actually detected by a virtual detector and give analytical expressions in both the real space and the momentum space. We prove that in the asymptotic limit when the time difference between wave packet creation and detection is large, resembling real detection processes, the obtained differential momentum distribution is equivalent to Fourier transforming the free-particle wave packets. We further consider the situation with the presence of laser fields.

\end{abstract}

\maketitle

\section{Introduction}

Interaction between a strong laser field and gas-phase atoms or molecules yields many important phenomena and applications, such as multiphoton and above-threshold ionization \cite{Agostini-79}, nonsequential double ionization \cite{Walker-94}, high-order harmonic generation \cite{McPherson-87, Ferray-88}, attosecond pulse generation\cite{Hentschel-01}, etc. In retrospect, it is kind of surprising to see that a combination of long familiar elements, namely, photons and atoms, could have led to so much new physics \cite{Brabec-RMP-00, Eberly-RMP-12} and pushed us into the unprecedented attosecond time domain \cite{Krausz-RMP-09}.

Theoretical understandings of the interaction between a strong laser field and an atomic or molecular target rely heavily on numerical wave function calculations due to the nonperturbative nature of the interaction. Such a calculation usually imposes a heavy computational load, especially when differential information is of interest, due to the need of a large numerical grid to keep the wave function, which may spread to hundreds to thousands of atomic units in space driven by the strong laser field.

Attempts have been made to reduce the computation load based on the observation that at large distances from the ion core, the effect of the Coulomb potential is weak, and the electron motion can be approximated as a free particle driven by the laser field. Such motion can be described analytically by a Volkov state \cite{Volkov-35} therefore further numerical integrations are not needed. Examples based on this approximation include a time-dependent surface flux method proposed by Scrinzi and coworkers \cite{Tao-12, Scrinzi-12}, and an analytical R-matrix method proposed by Smirnova and coworkers \cite{Torlina-12, Kaushal-13, Torlina-15}.

Another method, called a virtual detector (VD) method, proposed by Feuerstein and Thumm in the strong-field regime \cite{Thumm-03}, presents a radically different idea of approaching the same goal of reducing the computation load. Instead of seeking more economic ways of calculating or approximating the wave function evolving to the end of the laser pulse, the VD method extracts desired information (e.g., electron momentum distribution) from the wave function as early as possible, long before the end of the pulse, then destroys the wave function using for example an absorbing boundary. No large numerical grid is thus needed to keep the ever spreading wave function.

Based on the VD method, Wang et al. develop a hybrid quantum mechanical and classical trajectory approach that takes full account of the long range ion core Coulomb potential \cite{Wang-13}. In addition to being a tool of reducing the computation load, the VD method is recently found useful in helping to interpret strong-field ionization processes, especially the controversial tunneling ionization process. For example, Teeny et al. \cite{Teeny-16-PRL, Teeny-16-PRA} and Ni et al. \cite{Ni-16} use the VD method to extract tunnel-ionization-related information such as tunneling time, tunneling rate, position of tunneling exit, etc. Tian et al. \cite{Tian-17}  solve the controversy of electron longitudinal momentum at the tunneling exit \cite{Pfeiffer-12, Sun-14}.

In this paper we apply the VD method to free-particle wave packets. This seems to be a step back, since the VD method has been applied to more complicated situations involving the Coulomb potential, as cited above \cite{Wang-13,Teeny-16-PRL, Teeny-16-PRA, Ni-16, Tian-17}. However, we point out that previous studies have been mostly numerically oriented, and analytical insights are lacking. For example, it is conjectured in \cite{Thumm-03} that the differential momentum distributions of free-particle wave packets obtained using the VD method should be equivalent to Fourier transforming the wave packets, but no general prove is given except for the special case of a one-dimensional Gaussian wave packet. Such a general prove will be given in this paper. Therefore the current paper aims at providing more analytical insights into the VD method.

The importance of considering free-particle wave packets is reinforced considering that the electron is largely a free particle driven by the external laser field after being detached from the atom, as explained above. A deeper understanding on the virtual detection process of free-particle wave packets is essential for any further usage or development of the VD method.

This paper is organized as follows. In Section II we first introduce the basis idea of the VD method (II-A), then we explain what is actually detected by a VD, in both the real space (II-B) and the momentum space (II-C, II-D). We apply the VD method to general free-particle wave packets and prove that the differential momentum distribution obtained by the VD method is equivalent to Fourier transforming the wave packets (II-E, II-F). In Section III we use a 1D Gaussian wave packet to help the reader to better understand the VD method (III-A). Then we consider the effect of the presence of a laser field (III-B). A summary is given in Section IV.

\section{Theory and main results}

\subsection{Basic idea of the VD method}

A virtual detector is an imaginary device that locates at some fixed position in space and extracts information from the wave function passing through it. The information of interest usually includes the particle velocity or momentum (the meaning of which will be explained in detail later) and the probability current. Similarly, a real detector also locates at some fixed position in space and detects the probability current passing through it.

The differences between a virtual detector and a real detector are also obvious. First, a virtual detection process does not affect the wave function, whereas a real detection process collapses the wave function. Second, a virtual detector can be put anywhere in space, e.g. microscopically close to an interaction center, whereas a real detector always has a macroscopic distance from the interaction center. Third, a virtual detector does not have the realistic complications of a real detector, such as detection efficiency, random noises, response time, etc.

The original goal of proposing the VD method is to reduce the computation load of calculating differential momentum distributions in strong-field processes, such as molecular dissociation, atomic ionization, etc \cite{Thumm-03}. After information extraction by the VDs, further wave function evolution is not of concern so that it can be destroyed via an absorbing boundary. Therefore a large numerical grid is not needed to keep the whole wave function, which can spread to a distance of hundreds to thousands of atomic units.

\begin{figure}
  \centering
  \includegraphics[width=8cm]{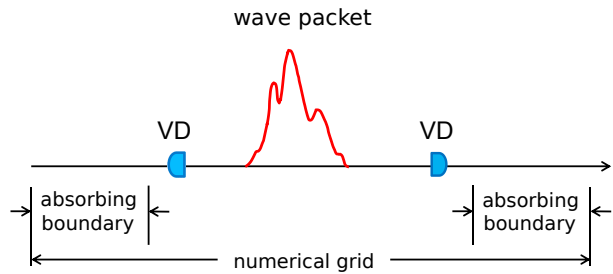}
  \caption{Illustration of the virtual detector method in one dimension. A wave packet is created near the center of the numerical grid. A VD is placed each side, extracting information from the wave packet. After passing through the VDs, the wave enters the absorbing boundary zone being absorbed.}\label{f.VD}
\end{figure}

A schematic illustration of the VD method is shown in Fig. \ref{f.VD} for one dimension, and extension to two or three dimensions is conceptually straightforward. A wave packet is created in the interaction center, via e.g. strong-field ionization, and it evolves as a function of time. A VD is placed on each side to extract information from the wave packet. After passing through the VDs, the wave packet enters the absorbing boundary zone being absorbed (destroyed) to avoid reflection.

\subsection{Flow velocity of the probability fluid}

A virtual detector extracts particle velocity (momentum) from the wave packet. This velocity is understood as the flow velocity of the probability fluid. If we write a wave function in the polar form
\beq \label{e.Psi_rt}
\Psi (\vec{r},t) = A(\vec{r},t) e^{i\phi(\vec{r},t)},
\eeq
with $A(\vec{r},t)$ and $\phi(\vec{r},t)$ real functions and $A(\vec{r},t) \ge 0$, then the probability density is
\beq \label{e.rho_rt}
\rho (\vec{r},t) = | \Psi(\vec{r},t) |^2 = A(\vec{r},t)^2,
\eeq
and the probability current can be shown as
\beqa
\vec{j}(\vec{r},t) &=& \frac{i\hbar}{2m} \left[ \Psi(\vec{r},t) \nabla \Psi^*(\vec{r},t) -c.c.  \right] \nonumber \\
&=& \rho (\vec{r},t) \frac{\hbar}{m} \nabla \phi (\vec{r},t).   \label{e.j_rt}
\eeqa

In analogy to the classical fluid equation $\vec{j}(\vec{r},t) = \rho (\vec{r},t) \vec{v} (\vec{r},t)$, one sees that the flow velocity is proportional to the spatial gradient of the phase of the wave function
\beqa
\vec{v} (\vec{r},t) &=& \frac{\hbar}{m} \nabla \phi (\vec{r},t), \text{  or} \label{e.flowvelocity}\\
\vec{p} (\vec{r},t) &=& \hbar \nabla \phi (\vec{r},t), \text{  or} \label{e.flowmomentum} \\
\vec{k} (\vec{r},t) &=& \nabla \phi (\vec{r},t).  \label{e.flowwavenumber}
\eeqa
These are the velocity/momentum/wave number extracted by a VD located at position $\vec{r}$ at time $t$. We emphasize that it is legal and meaningful to talk about the flow velocity at a fixed position, and the VD method is therefore closely related to the field of quantum hydrodynamics \cite{Wyatt-book} and trajectory formalisms of quantum mechanics \cite{Bohm-1, Bohm-2, Madelung, Takabayasi}.

Another way of obtaining the flow velocity is to apply the momentum operator to the wave function
\beqa
\hat{p} \Psi (\vec{r},t) &=& -i \hbar \nabla \Psi (\vec{r},t) \nonumber \\
&=& \left[ \hbar \nabla \phi (\vec{r},t) - i \hbar \frac{\nabla A (\vec{r},t)}{A(\vec{r},t)} \right] \Psi(\vec{r},t),  \label{e.eigen_p}
\eeqa
which yields a complex ``eigenvalue". The velocity associated to the real part is exactly the flow velocity of Eq. (\ref{e.flowvelocity}), and the velocity associated to the imaginary part is called the Einstein osmotic velocity, related to quantum diffusions.

\subsection{Flow velocity in the momentum space}

To have a better understanding about the flow velocity of the probability fluid, we give its expression in the momentum space. Assume a wave packet is created at time $t = 0$, when it can be written as
\beq
\Psi (\vec{r},t=0) = \frac{1}{(2\pi)^{3/2}} \int \tilde{\psi} (\vec{k}) e^{i \vec{k} \cdot \vec{r}} d^3 \vec{k},
\eeq
where $\tilde{\psi} (\vec{k})$ is the complex Fourier component. Then the wave packet evolves with time according to
\beq \label{e.Psi_rt_FT}
\Psi (\vec{r},t) = \frac{1}{(2\pi)^{3/2}} \int \tilde{\psi} (\vec{k}) e^{i \Delta(\vec{k},\vec{r},t)} d^3 \vec{k},
\eeq
where we have defined
\beq   \label{e.Delta}
\Delta(\vec{k},\vec{r},t) \equiv \vec{k} \cdot \vec{r} - \om (\vec{k})t
\eeq
with $\om (\vec{k})$ the dispersion relation.

From Eqs. (\ref{e.Psi_rt}) and (\ref{e.Psi_rt_FT}), we can get
\beq
e^{i\phi(\vec{r},t)} = \frac{1}{A(\vec{r},t)} \frac{1}{(2\pi)^{3/2}} \int \tilde{\psi} (\vec{k}) e^{i \Delta(\vec{k},\vec{r},t)} d^3 \vec{k}.
\eeq
Take logarithms of both sides and we get
\beqa
i\phi(\vec{r},t) = \ln \left\{ \int \tilde{\psi} (\vec{k}) e^{i \Delta(\vec{k},\vec{r},t)} d^3 \vec{k} \right\} \nonumber \\
- \ln A(\vec{r},t) - \ln (2\pi)^{3/2}.
\eeqa
Then take the spatial gradient of the phase function
\beq
\nabla \phi(\vec{r},t) = \frac{\int \vec{k} \tilde{\psi} (\vec{k}) e^{i \Delta(\vec{k},\vec{r},t)} d^3 \vec{k}}{\int \tilde{\psi} (\vec{k}) e^{i \Delta(\vec{k},\vec{r},t)} d^3 \vec{k}} + i\frac{\nabla A(\vec{r},t)}{A(\vec{r},t)}.
\eeq
Because $\phi(\vec{r},t)$ and $A(\vec{r},t)$ are both real functions, we conclude that the imaginary part of the first term on the right hand side must cancel the second term (the proof of this point can be done easily which will not be shown here). Then we have
\beq \label{e.gradphi}
\nabla \phi(\vec{r},t) = \operatorname{Re} \left\{ \frac{\int \vec{k} \tilde{\psi} (\vec{k}) e^{i \Delta(\vec{k},\vec{r},t)} d^3 \vec{k}}{\int \tilde{\psi} (\vec{k}) e^{i \Delta(\vec{k},\vec{r},t)} d^3 \vec{k}} \right\}.
\eeq
We see that formally $\nabla \phi(\vec{r},t)$ is the real part of the expectation value of $\vec{k}$, consistent to the interpretation given by Eq. (\ref{e.eigen_p}) in the real space.

\subsection{Probability current in the momentum space}

From Eqs. (\ref{e.rho_rt}) and (\ref{e.Psi_rt_FT}), the probability density can be written in the form
\beqa \label{e.rho_rt_FT}
\rho (\vec{r},t) = \frac{1}{(2\pi)^{3}} \int \tilde{\psi} (\vec{k}) e^{i \Delta(\vec{k},\vec{r},t)} d^3 \vec{k}  \nonumber \\
\times \int \tilde{\psi}^* (\vec{k}) e^{-i \Delta(\vec{k},\vec{r},t)} d^3 \vec{k}.
\eeqa

From Eqs. (\ref{e.j_rt}), (\ref{e.gradphi}), and (\ref{e.rho_rt_FT}), the probability current can be shown as
\beqa
\vec{j}(\vec{r},t) = \frac{1}{(2\pi)^{3}} \frac{\hbar}{m} \operatorname{Re} \Bigg\{ \int \vec{k} \tilde{\psi} (\vec{k}) e^{i \Delta(\vec{k},\vec{r},t)} d^3 \vec{k} \nonumber \\
\times \int \tilde{\psi}^* (\vec{k}) e^{-i \Delta(\vec{k},\vec{r},t)} d^3 \vec{k}  \Bigg\}.  \label{e.j_rt_FT}
\eeqa

We see that further evaluations of the flow velocity and the probability current involve the following two integrals
\beqa
I_1 &=& \int \tilde{\psi} (\vec{k}) e^{i \Delta(\vec{k},\vec{r},t)} d^3 \vec{k},  \label{e.int1}\\
I_2 &=& \int \vec{k} \tilde{\psi} (\vec{k}) e^{i \Delta(\vec{k},\vec{r},t)} d^3 \vec{k}. \label{e.int2}
\eeqa

\subsection{Free-particle wave packets and the asymptotic limit}

The above formalism is general and exact. No approximations have been made up to now. Next let us consider a free-particle wave packet by specifying the dispersion relation
\beq
\om (\vec{k}) = \frac{\hbar k^2}{2m},  \label{e.omega_k}
\eeq
which makes the phase given in Eq. (\ref{e.Delta}) quadratic in $\vec{k}$. Completing the square, one gets
\beq  \label{e.phase}
\Delta(\vec{k},\vec{r},t) = -\frac{\hbar t}{2m} \left( \vec{k} -\frac{m \vec{r}}{\hbar t} \right)^2 + \frac{m r^2}{2\hbar t}.
\eeq

In the asymptotic limit where $t$ is large, the term $ e^{i \Delta (\vec{k},\vec{r},t)}$ is highly oscillatory except in the vicinity of the stationary wave vector
\beq
\vec{k}_s = \frac{m \vec{r}}{\hbar t}. \label{e.ks}
\eeq
Recall that $\vec{r}$ is the location of the virtual detector. In this asymptotic limit $I_1$ and $I_2$ can be evaluated simply
\beqa
I_1 &=& \tilde{\psi} (\vec{k}_s) \exp \left( i \frac{m r^2}{2\hbar t} \right) \left( \frac{2\pi m}{\hbar t} \right)^{3/2} e^{-i\frac{3\pi}{4}}, \\
I_2 &=& \vec{k}_s I_1.
\eeqa
We have used the Fresnel integral
\beq
\int e^{\pm ibx^2} d^3 \vec{x} = \left( \frac{\pi}{b} \right)^{3/2} e^{\pm i \frac{3\pi}{4}},  \ \text{for } b>0.
\eeq
The asymptotic limit and the stationary phase condition effectively disentangle the wave vector superpositions appearing in Eqs. (\ref{e.int1}) and (\ref{e.int2}).

Using Eq. (\ref{e.gradphi}) the flow velocity becomes
\beq
\vec{v} (\vec{r},t) = \frac{\hbar}{m} \nabla \phi(\vec{r},t) = \frac{\hbar}{m} \operatorname{Re} \left( \frac{I_2}{I_1} \right) = \frac{\hbar \vec{k}_s}{m} = \frac{\vec{r}}{t},
\eeq
which is just the classical velocity for the particle flying from the interaction center (where the wave packet is created) to the virtual detector.

Using Eq. (\ref{e.j_rt_FT}) the probability current becomes
\beqa
\vec{j}(\vec{r},t) &=& \frac{1}{(2\pi)^{3}} \frac{\hbar}{m} \operatorname{Re} \left( I_1^* I_2 \right) \nonumber \\
&=& \left( \frac{m}{\hbar} \right)^3 \frac{\vec{r}}{t^4} \left| \tilde{\psi} \left( \frac{m\vec{r}}{\hbar t} \right) \right|^2   \nonumber \\
&=& \left( \frac{m}{\hbar} \right)^2 \frac{1}{t^3} \vec{k}_s \left| \tilde{\psi} \left( \vec{k}_s \right) \right|^2.   \label{e.j_rt_asymp}
\eeqa
This formula links the probability current detected by a VD at position $\vec{r}$ and time $t$, with the Fourier component $|\tilde{\psi} (\vec{k}_s)|^2$ of the wave packet. It provides a practical way of obtaining the Fourier components of free-particle wave packets.

\subsection{The differential momentum distribution}

A virtual detector, located at position $\vec{r}$, detects at each time $t$ the flow momentum of the wave passing through it, as well as the probability current, which is used as the weight of the corresponding momentum. By accumulating the momentum and the corresponding weight over time, one obtains the differential momentum distribution. We now show that the momentum distribution obtained this way, a scheme that resembles the real experiments, is equivalent to Fourier transforming the free-particle wave packet.

The total probability of particle detection is
\beq
\int_{0}^{\infty} \left[\vec{j}(\vec{r},t) \cdot d\vec{S} \right] dt
= \int \left[\vec{j}(\vec{r},\vec{k}_s) \cdot d\vec{S} \right] \left| \frac{dt}{d^3 \vec{k}_s} \right| d^3 \vec{k}_s,
\eeq
where $d\vec{S} = \hat{r} dS = \hat{r} r^2 d\Omega$ is the area element of the detector, with $d\Omega$ the spanned solid angle. Therefore the probability density of finding a particle within a small momentum volume $d^3 \vec{k}_s$ around $\vec{k}_s$ is
\beqa
P(\vec{k}_s) &=& \left[\vec{j}(\vec{r},\vec{k}_s) \cdot d\vec{S} \right] \left| \frac{dt}{d^3 \vec{k}_s} \right|         \nonumber \\
&=& j (\vec{r},\vec{k}_s) r^2 d\Omega \left| \frac{dt}{k_s^2 dk_s d\Omega} \right|.  \label{e.Pks}
\eeqa

Using the stationary phase relation given by Eq. (\ref{e.ks}), we have
\beq
\frac{dt}{dk_s} = - \frac{m r}{\hbar k_s^2}.
\eeq
Then the differential momentum distribution of Eq. (\ref{e.Pks}), after a few straightforward algebra steps, becomes precisely
\beq
P(\vec{k}_s) = \left| \tilde{\psi} (\vec{k}_s) \right|^2.
\eeq
We have finished our proof that in the asymptotic limit, the differential momentum distribution obtained by the VD method is equivalent to Fourier transforming the free-particle wave packet. This is the main result of this paper.

\section{Examples and discussions}

\subsection{One-dimensional Gaussian wave packet}

In this subsection we use a 1D Gaussian wave packet to illustrate the flow velocity, the probability current, and the asymptotic limit. The purpose is to help the reader to better understand these physical quantities. For the sake of convenience, we consider an electronic wave packet and use atomic units, so $\hbar = m = 1$. (Note that atomic units are used only in the current subsection III-A)

Consider a Gaussian wave packet with spatial width $\Gamma$ at time $t=0$
\beq
\Psi (x,t=0) = \frac{1}{\sqrt{\Gamma} \pi^{1/4}} \exp \left( -\frac{x^2}{2\Gamma^2} \right) \exp (i k_0 x),
\eeq
which centers at $x=0$ and moves with group velocity $k_0$. The corresponding Fourier transform is
\beq
\tilde{\psi} (k) = \frac{\sqrt{\Gamma}}{\pi^{1/4}} \exp \left[ -\frac{\Gamma^2}{2} (k-k_0)^2 \right].
\eeq

Using the 1D version of Eqs. (\ref{e.gradphi}), (\ref{e.int1}), (\ref{e.int2}), and (\ref{e.omega_k}), we easily get
\beqa
k(x,t) &=& \frac{\partial \phi(x,t)}{\partial x} = \operatorname{Re} \left( \frac{I_2}{I_1}  \right)  \nonumber \\
&=& \frac{\Gamma^4 k_0 + x t}{\Gamma^4 + t^2},  \label{e.kxt_Gaussian}
\eeqa
which is the same as Eq. (14) of Ref. \cite{Thumm-03} using a different approach. Note that this expression is valid for all $t$ since the asymptotic limit has not been applied. We see that $k(x,t=0) = k_0$ and $k(x,t \rightarrow \infty) = x/t$, the classically expected velocity. We also see that the asymptotic limit is fulfilled when the following two conditions are met: $xt \gg \Gamma^4 k_0$ and $t^2 \gg \Gamma^4$. Or we can combine them as $t \gg \max \{ \Gamma^2, \Gamma^4 k_0/x \}$.

\begin{figure} [t!]
  \centering
  \includegraphics[width=4.2cm]{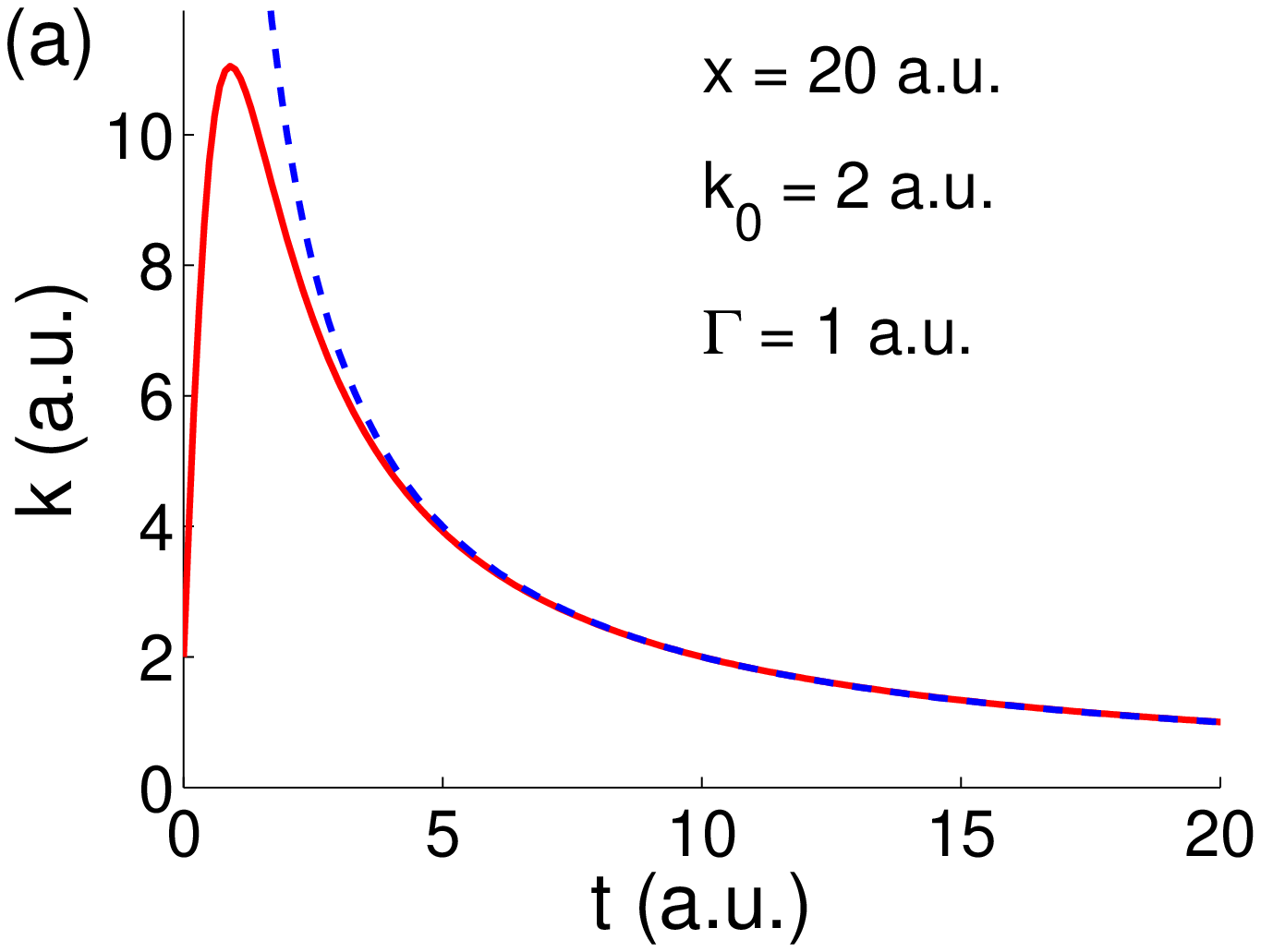}
  \includegraphics[width=4.2cm]{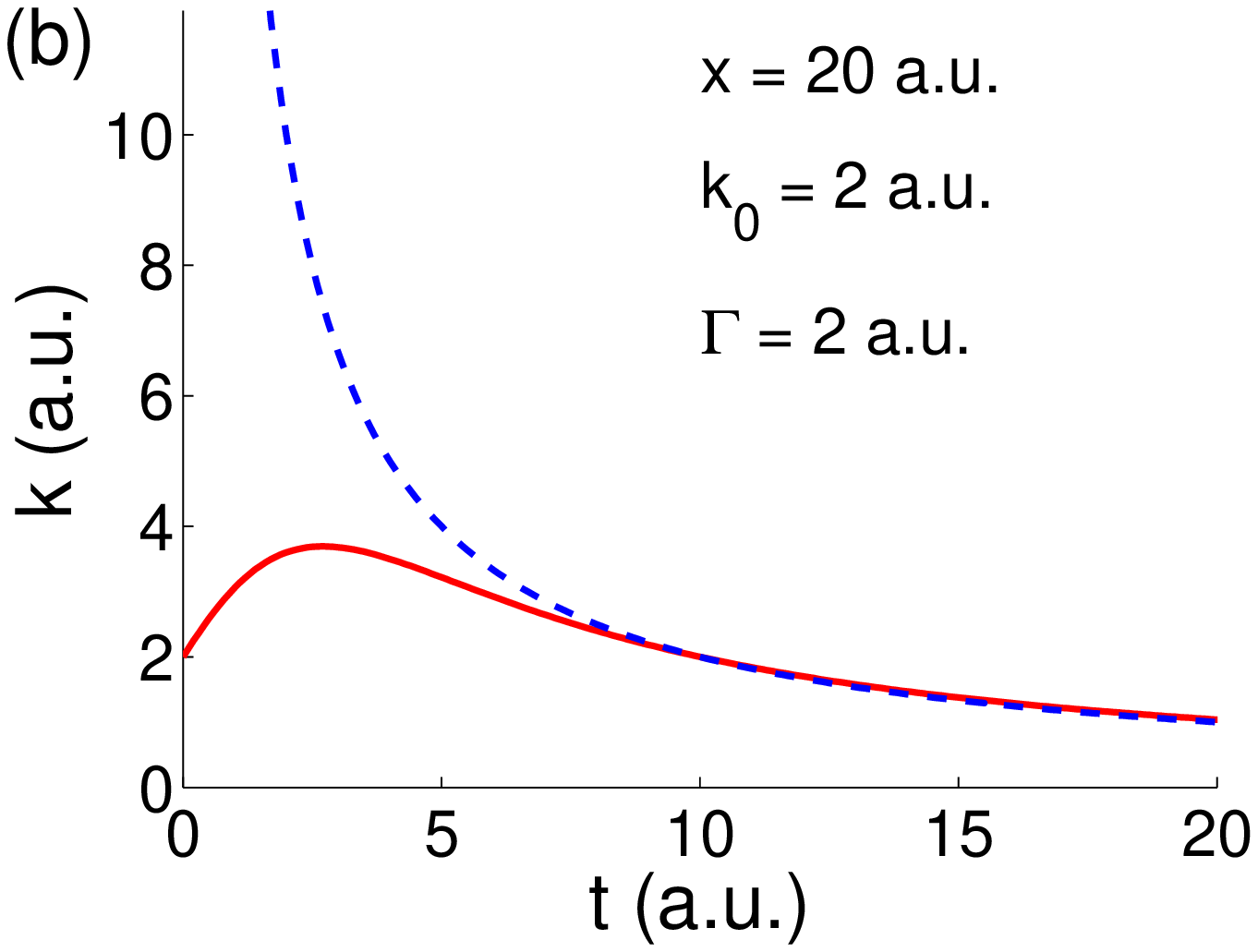}
  \caption{Velocity detected by a VD located at $x = 20$ a.u. The Gaussian wave packet is moving with a group speed $k_0 = 2$ a.u. The initial width of the wave packet is $\Gamma = 1$ a.u. (a) and $\Gamma = 2$ a.u. (b). The red solid curve in each panel is the velocity detected by the VD, given by Eq. (\ref{e.kxt_Gaussian}). The blue dashed curve in each panel is the classically expected velocity $x/t$.}\label{f.kxt}
\end{figure}

\begin{figure} [t!]
  \centering
  \includegraphics[width=4.2cm]{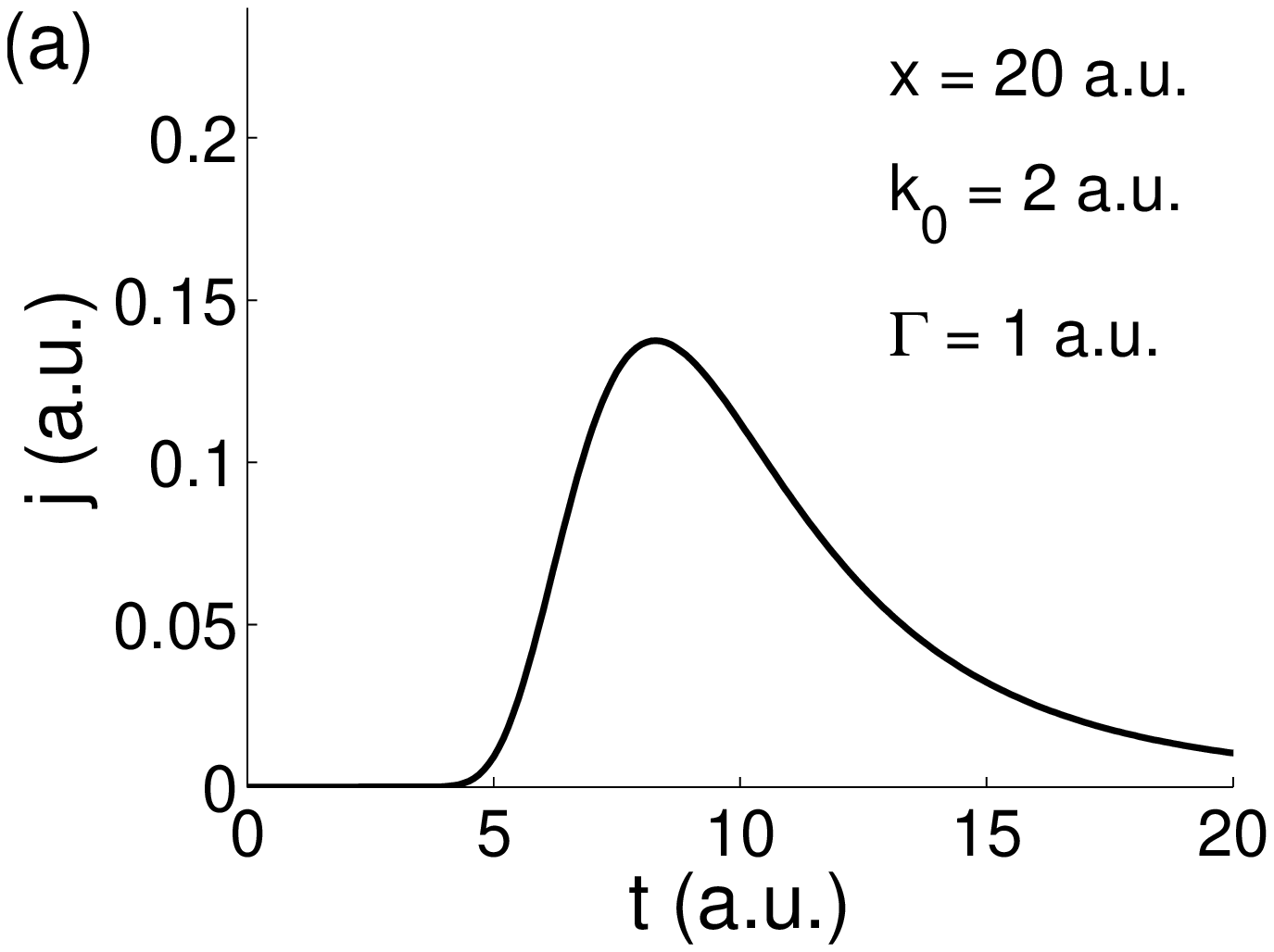}
  \includegraphics[width=4.2cm]{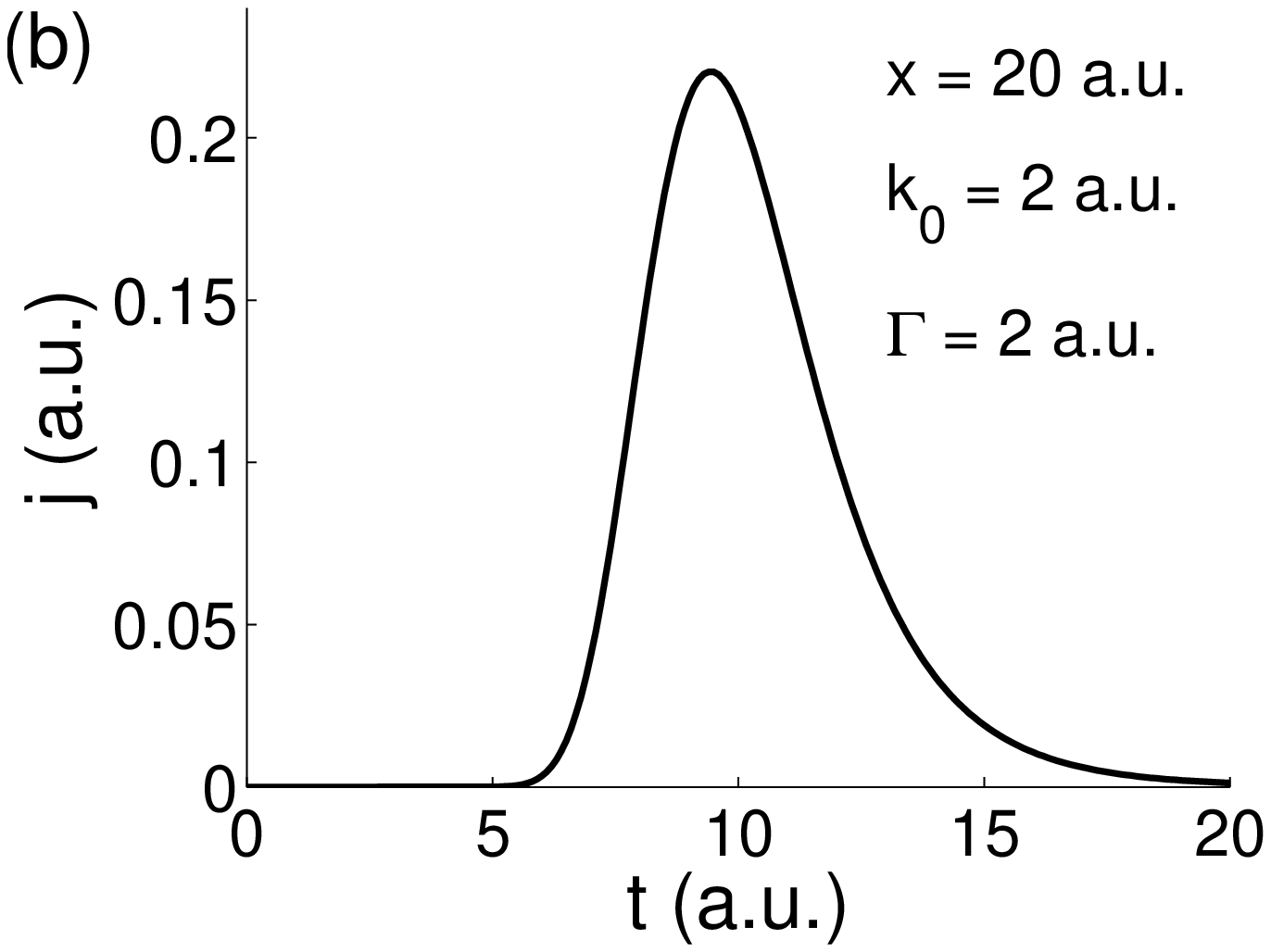}
  \caption{Probability current detected by a VD located at $x = 20$ a.u. as a function of time, corresponding to the same two cases as Fig. \ref{f.kxt}.}\label{f.jxt}
\end{figure}

An illustration of $k(x,t)$ is given in Fig. \ref{f.kxt} for a fixed VD position $x=20$ a.u. The parameters of the wave packet are chosen as $k_0 = 2$ a.u., $\Gamma = 1$ a.u. (panel a) and $\Gamma = 2$ a.u. (panel b). The red (solid) curve in each panel is the velocity detected by the VD as a function of time, viz. $k(x,t)$ given by Eq. (\ref{e.kxt_Gaussian}). The blue (dashed) curve in each panel is the classically expected velocity, given by $x/t$. When the asymptotic condition is met, the velocity detected by the VD approaches the classical velocity. One can check that the asymptotic condition for each case is $t \gg 1$ a.u. and $t \gg 4$ a.u., respectively. It is worth mentioning that in the asymptotic limit the time $t$ does not have to be macroscopic.

For $t=0$, $k = k_0$ and is independent of the position of the VD. Before reaching the asymptotic limit where a single Fourier component dominates due to the stationary-phase condition, the velocity detected by the VD, according to Eq. (\ref{e.gradphi}), is the result of the superposition of all Fourier components.

Using the 1D version of Eq. (\ref{e.j_rt_FT}) the probability current can be obtained simply
\beqa
j(x,t) &=& \frac{1}{2\pi} \frac{\hbar}{m} \operatorname{Re} (I_1^* I_2)  \nonumber \\
&=& \frac{\Gamma k(x,t)}{\sqrt{\pi (\Gamma^4 + t^2)}} \exp \left[ -\frac{\Gamma^2 (x-k_0 t)^2}{\Gamma^4 + t^2} \right],   \label{e.jxt_Gaussian}
\eeqa
where $k(x,t)$ is given by Eq. (\ref{e.kxt_Gaussian}). This result is the same as Eq. (16) of Ref. \cite{Thumm-03}. Fig. \ref{f.jxt} shows the probability current passing through the VD corresponding to the two cases of Fig. \ref{f.kxt}. Since the VD locates at $x = 20$ a.u. and the wave packet moves with a group velocity of $k_0 = 2$ a.u., one expects a peak probability current around $t = 10$ a.u. Because the wave packet in the first case (with $\Gamma = 1$ a.u.) is initially more localized in space than the second case (with $\Gamma = 2$ a.u.), we expect the first wave packet to spread more quickly than the second one. Therefore the probability current shown in panel (a) starts earlier and ends later in time than the one shown in panel (b).

\subsection{Free-particle wave packet in a laser field}

In a strong-field experiment the interaction between the laser field and the atomic or molecular target happens in the laser focus. Interaction resultants, such as emitted electrons, ionized atoms, molecular fragments, etc., fly out from the laser focus to the detector, which has a macroscopic distance from the laser focus. The detection happens long after ($\sim$ nanoseconds) the laser pulse ($\sim$ femtoseconds) is over. In fact, during a laser pulse of a few tens of femtoseconds, the resultants are deep inside the focus, having no chance of experiencing the spatial gradient of the laser focus before the laser pulse is over.

A virtual detector, in contrast, has the flexibility of putting anywhere in space, including places inside the laser focus. For the VD method being useful in reducing the computational load of strong-field processes, the VDs have to be put inside the laser focus, with distances from the interaction center much smaller than the length of the numerical grid keeping the whole wave function. Therefore it is important to understand what is detected by a VD inside a laser field.

We consider here the simplest situation that a free-particle wave packet is created at time $t=0$ inside a laser field (which may not be zero then). Then the theory and formalism given in Section II apply to the current situation, provided that some modifications are made. Specifically, the dispersion relation in Eq. (\ref{e.omega_k}) needs to be modified to
\beq \label{e.omega_k_laser}
\om (\vec{k},t) = \frac{1}{2\hbar m} \left[ \hbar \vec{k} - q \mathcal{\vec{A}} (t) \right]^2,
\eeq
where $q$ is the charge of the particle and $\mathcal{\vec{A}} (t)$ is the vector potential of the laser field. The spatial dependence of the vector potential is not considered due to the reason just explained above.

The phase of Eq. (\ref{e.phase}) becomes (with a subscript $L$ denoting for laser)
\beqa
\Delta_L (\vec{k},\vec{r},t) &=& \vec{k} \cdot \vec{r} - \int_{0}^{t} \om (\vec{k},t') dt'  \nonumber \\
&=& \vec{k} \cdot \vec{r} - \frac{\hbar k^2 t}{2m} + \vec{k} \cdot \frac{q}{m} \int_{0}^{t} \mathcal{\vec{A}} (t') dt' \nonumber \\
& &  - \frac{q^2}{2\hbar m} \int_{0}^{t} \mathcal{A}^2 (t') dt'  \nonumber \\
&\equiv& \vec{k} \cdot \vec{r} - \frac{\hbar k^2 t}{2m} + \vec{k} \cdot \vec{l} (t) + \xi(t).     \label{e.phase_laser}
\eeqa
Two short hand notations $\vec{l} (t)$ and $\xi (t)$ are introduced, whose definitions can be seen comparing the third step with the second step. The first two terms on the right hand side are the same as those of the no-laser situation, and the additional two terms are introduced by the laser field. $\vec{l} (t)$ has the units of length and is the additional displacement induced by the laser electric field from time $0$ to $t$. $\xi (t)$ is an additional phase.

In the asymptotic limit when $t$ is large, the phase is highly oscillatory except in the vicinity of the stationary $\vec{k}$ vector, which, by requiring $\partial \Delta_L / \partial \vec{k} = 0$, is now
\beq
\vec{k}_s = \frac{m}{\hbar t} \left[ \vec{r} + \vec{l} (t) \right].  \label{e.ks_laser}
\eeq
We see that the laser-introduced additional displacement $\vec{l} (t)$ enters into the picture. As explained above, and demonstrated in Fig. \ref{f.kxt}, fulfilling the asymptotic condition does not necessarily conflict the configuration that the VDs are put inside the laser focus and close to the interaction center.

Following similar steps as in Section II-D and II-E, we get the probability current detected by a VD located at $\vec{r}$ and at time $t$
\beq  \label{e.j_rt_laser}
\vec{j} (\vec{r},t) = \left( \frac{m}{\hbar} \right)^2 \frac{1}{t^3} \left[ \vec{k}_s - \frac{q}{\hbar} \mathcal{\vec{A}} (t) \right] \left| \tilde{\psi}(\vec{k}_s) \right|^2,
\eeq
which reduces to Eq. (\ref{e.j_rt_asymp}) as the laser field vanishes.

Eq. (\ref{e.j_rt_laser}), together with Eq. (\ref{e.ks_laser}), tells us how to decode the Fourier components of the free-particle wave packet from the probability current detected by the VDs, in the presence of a laser electric field.

\section{Summary}

We apply the recently emerging virtual detector method to general free-particle wave packets. Although the VD method has been used in several previous studies \cite{Thumm-03, Wang-13, Teeny-16-PRL, Teeny-16-PRA, Ni-16, Tian-17}, analytical insights about the method itself are lacking. Applying the method to free-particle wave packets helps to obtain such insights.

A virtual detector is an imaginary device placed at a fixed position in space and extracts information from the wave passing through it. Information of interest includes (but not limits to) the particle velocity and the probability current. The particle velocity is understood as the flow velocity of the probability fluid. The probability current is the weight of the corresponding velocity. By accumulating the velocity/momentum and the corresponding weight over time, differential momentum distribution of the particle can be obtained. The main result of the current paper is a proof that the momentum distribution obtained by the VD method, a scheme resembling real experiments, is equivalent to the normal method of Fourier transforming the wave packet.

We also consider the situation of virtual detecting a free-particle wave packet inside a laser field. We show that the asymptotic velocity determined by the stationary phase condition, as well as the probability current, will be modified by the laser electric field. Fourier components of the wave packet, however, can still be obtained from the detected probability current.

Acknowledgement: We acknowledge discussions with Prof. J. H. Eberly. This work was supported by China NSF No. 11774323 and China Science Challenge Project No. TZ2017005.

\end{document}